\documentclass[letter]{aa}

\usepackage{amsmath}
\usepackage{graphicx,color}
\usepackage{natbib}
\usepackage{url}
\usepackage[varg]{txfonts}
\usepackage{threeparttable}
\usepackage{booktabs, makecell}

\def\ero{\textit{eROSITA}\xspace}

\def\gai{\textit{Gaia}\xspace}

\def\ros{\textit{ROSAT}\xspace}
\def\exo{\textit{EXOSAT}\xspace}

\def\swi{\textit{Swift}/XRT\xspace}

\def\xmmn{\textit{XMM-Newton}\xspace}

\newcommand{\be}{\begin{equation}}
\newcommand{\ee}{\end{equation}}

\newcommand\att{{eRASSt\,J192932.9$-$560346}\xspace}

\newcommand\fergs{\ensuremath{\mathrm{erg}\,\mathrm{cm}^{-2}\,\mathrm{s}^{-1}}\xspace}
\newcommand\lxu{\ensuremath{\mathrm{erg}\,\mathrm{s}^{-1}}\xspace}

\newcommand\kmps{\ensuremath{\mathrm{km}\,\mathrm{s}^{-1}}\xspace}
\newcommand\msun{\ensuremath{M_\odot}\xspace}

\newcommand\rat{\ensuremath{\mathrm{s}^{-1}}\xspace}

\begin{document}

\title{Discovery of \att: a bright, two-pole accreting, eclipsing polar}

\author{Axel Schwope\inst{1}
\and David A.H.~Buckley \inst{2, 3}
\and Adam Malyali\inst{4}
\and Stephen Potter\inst{2,5}
\and Ole König\inst{6}
\and Riccardo Arcodia\inst{4}
\and Mariusz Gromadzki\inst{7}
\and Arne Rau\inst{4}
}

\institute{Leibniz-Institut f\"ur Astrophysik Potsdam (AIP), An der Sternwarte 16, 14482 Potsdam, Germany\\
\email{aschwope@aip.de}
\and
South African Astronomical Observatory, P.O. Box 9, Observatory, 7935, Cape Town, South Africa
\and
Department of Astronomy, University of Cape Town, Private Bag X3, Rondebosch 7701, South Africa
\and 
     Max-Planck-Institut f\"ur extraterrestrische Physik,
     Gie{\ss}enbachstra{\ss}e, 85748 Garching, Germany
\and
Department of Physics, University of Johannesburg, PO Box 524, Auckland Park 2006, South Africa
\and
Dr.~Karl Remeis-Sternwarte \& Erlangen Centre for Astroparticle Physics, Friedrich-Alexander-Universit\"at Erlangen-N\"urnberg, Sternwartstr.~7, 96049 Bamberg, Germany
\and
Astronomical Observatory, University of Warsaw, Al. Ujazdowskie 4, 00-478 Warszawa, Poland
}

\authorrunning{Schwope et al.}
\titlerunning{Discovery of \att, a bright, eclipsing polar}
\date{\today}


\abstract
{We report the discovery of a bright ($V\sim15$), eclipsing, two-pole accreting magnetic cataclysmic variable (CV), a polar, as counterpart of the \ero and \gai transients \att and Gaia21bxo. Frequent large amplitude changes of its brightness at X-ray and optical wavelengths by more than 4 magnitudes was indicative of a CV nature of the source. Identification spectra obtained with the 10m SALT telescope revealed the typical features of a magnetic CV, strong, broad He{\sc I}, He{\sc II} and hydrogen Balmer emission lines superposed on a blue continuum. Time-resolved photoelectric polarimetry revealed circular polarization to vary from -20\% to +20\%, and linear polarization from 0\% to 10\% confirming the system to be magnetic CV of the polar subclass. High cadence photometry revealed deep, structured eclipses, indicating that the system is a two pole accretor. The orbital period determined from the eclipse times is $92.5094\pm0.0002$ min. The X-ray spectrum is thermal only and the implied luminosity is $L_{\rm X}=2.2\times10^{31}$\,\lxu at the \gai-determined distance of 376\,pc.
}

\maketitle

\section{Introduction}
The \ero instrument \citep{predehl+21} onboard the Spektrum-Roentgen-Gamma spacecraft \citep[SRG; ][]{sunyaev+21} scans the X-ray sky with an imaging telescope in the energy band between 0.2 and 10 keV since December 2019. Eventually it shall performs eight independent all-sky surveys, called eRASSn (n=1\dots8), each lasting half a year. At the time of writing eRASS3 is ongoing. 

Given the sensitivity of \ero, several thousand new compact accreting binaries were expected to be found in the final stacked version of the \ero all-sky survey called eRASS:8 \citep[][]{merloni+12, schwope12}. Their identification among the several million point-like X-ray sources detected with the sensitive instrument needs comprehensive optical identification programmes that have just begun within the SDSS or are planned with the 4MOST facility \citep{kollmeier+17,dejong+19}. Before arrival of the stacked \ero catalogues, new identifications are posssible by exploiting the multi-dimensional parameter space spanned by time variability (at X-ray and other wavelengths), optical, infrared and X-ray colors, location in the color-magnitude diagram and similar. 

Cataclysmic variables are semi-detached close compact binaries containing a Roche-lobe filling donor star and an accreting white dwarf. While accretion in non-magnetic CVs happens via disks that act as mass storage, the accretion processes are much more direct in the magnetic systems that lack a disk as intermediary. Hence, changes in the mass transfer rate from the donor to the white dwarf may manifest themselves instantaneously in the observed brightness of a source. Two subclasses are defined among the magnetic CVs, the polars, which typically have short orbital periods ($P_{\rm orb} < 2$ hours) and similar rotation periods of the white dwarfs within a few percent, and the intermediate polars, which typically are long-period systems ($P_{\rm orb} > 3$ hours) and posses a more or less freely spinning white dwarf. The two classes are of prime interest for close binary evolution and the formation of strong magnetic fields in white dwarfs \citep[e.g.][]{schreiber+21}. As a group they are relevant for the total X-ray output of our own Milky Way galaxy \citep{revnivtsev+09}. 

Here we report the discovery of a bright, eclipsing, magnetic CV as counterpart of a variable eRASS-source. It attracted our attention due to its pronounced X-ray variability between the first three all-sky X-ray scans. The new \ero source was associated with a nearby \gai object (offset position 2\farcs6) which was noticed earlier by the \gai alert team to be a possible CV given its extreme photometric variability \citep{hodgkin+21}. 

We describe the initial \ero results (Sect.~\ref{s:ero}), put the data in context with archival X-ray observations (Sect.~\ref{s:xarc}), report our SALT identification spectra toghether with time-resolved photometry and polarimetry obtained from SAAO (Sect.~\ref{s:fup}), which alltogether uniquely identify the source. It offers great potential for in-depth studies of the accretion geometry via detailed eclipse analysis, comparable to \exo and \ros discoveries made some 30 years ago \citep[see e.g.][for UZ For and HU Aqr]{beuermann+88, schwope+93}.

\section{X-ray observations with \ero \label{s:ero}}

During eRASS3, \ero re-detected a source 
at $F_{\rm X} (0.6-2.3 {\rm keV}) = (2.6\pm0.4)\times 10^{-12}$\,\fergs, which was undetected in eRASS2 ($F_{\rm X}(0.6-2.3 {\rm keV}) < 8 \times 10^{-14}$\,\fergs) but previously found at $F_{\rm X}(0.6-2.3 {\rm keV}) = (1.8\pm0.3)\times 10^{-12}$\,\fergs\ in eRASS1. The X-ray source position determined in eRASS3 is RA(J2000)=19:29:32.9, DEC(J2000)=-56:03:46 with a statistical uncertainty of 1.3 arcsec. 

\begin{figure}[t]
\resizebox{\hsize}{!}{\includegraphics[clip=,angle=-90]{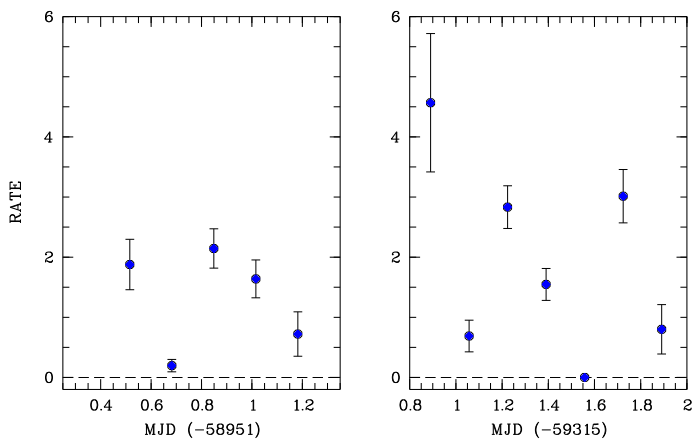}}
\caption{Full-band eRASS1 and eRASS3 X-ray light curves (0.2-10 keV). Shown are the mean rates per eroday \label{f:elcs}}
\end{figure}

\begin{figure}[t]
\resizebox{\hsize}{!}{\includegraphics{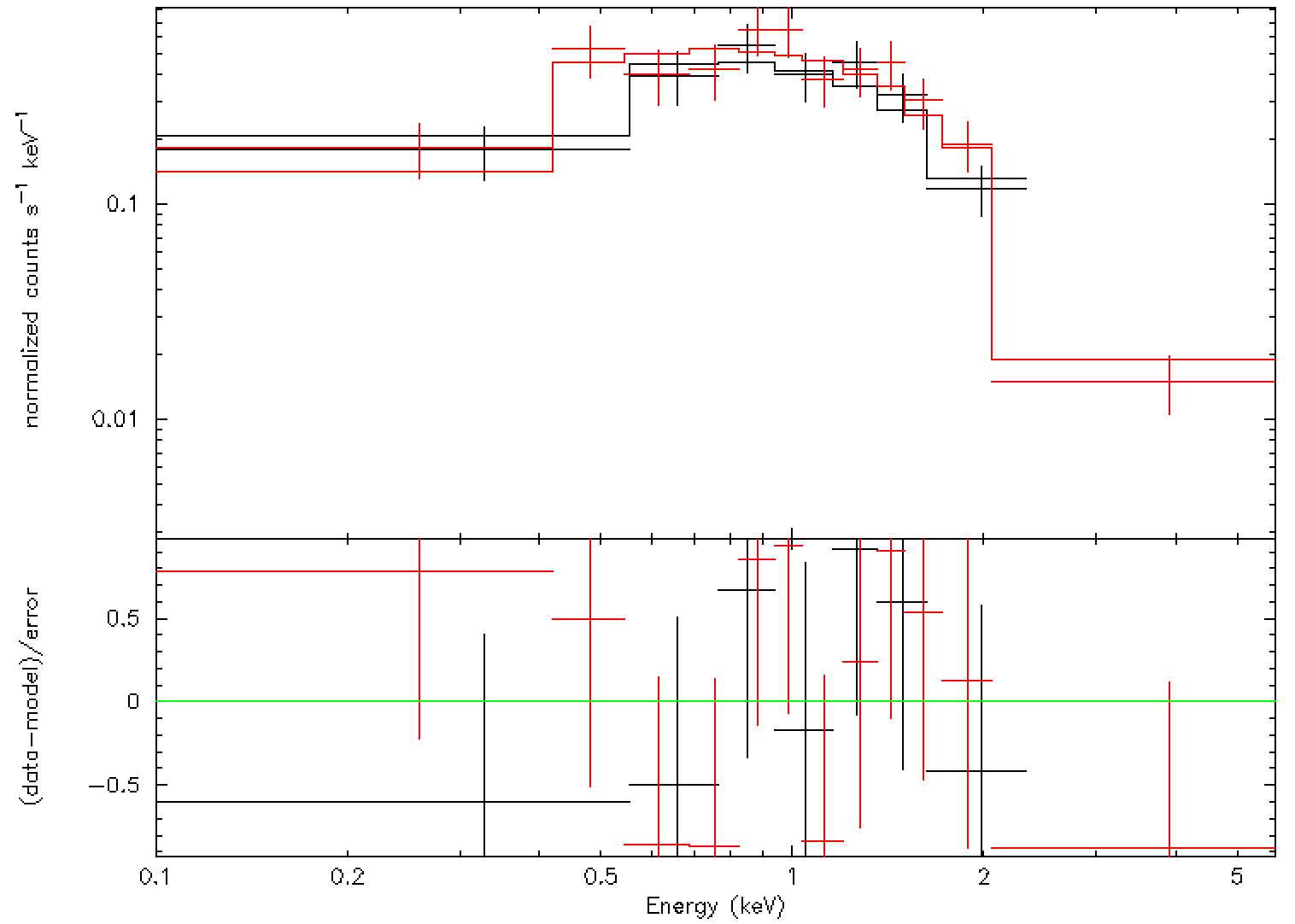}}
\caption{eRASS1 (black) and eRASS3 (red) spectra fitted with a thermal model \label{f:espc}}
\end{figure}

\begin{figure}[t]
\resizebox{\hsize}{!}{\includegraphics{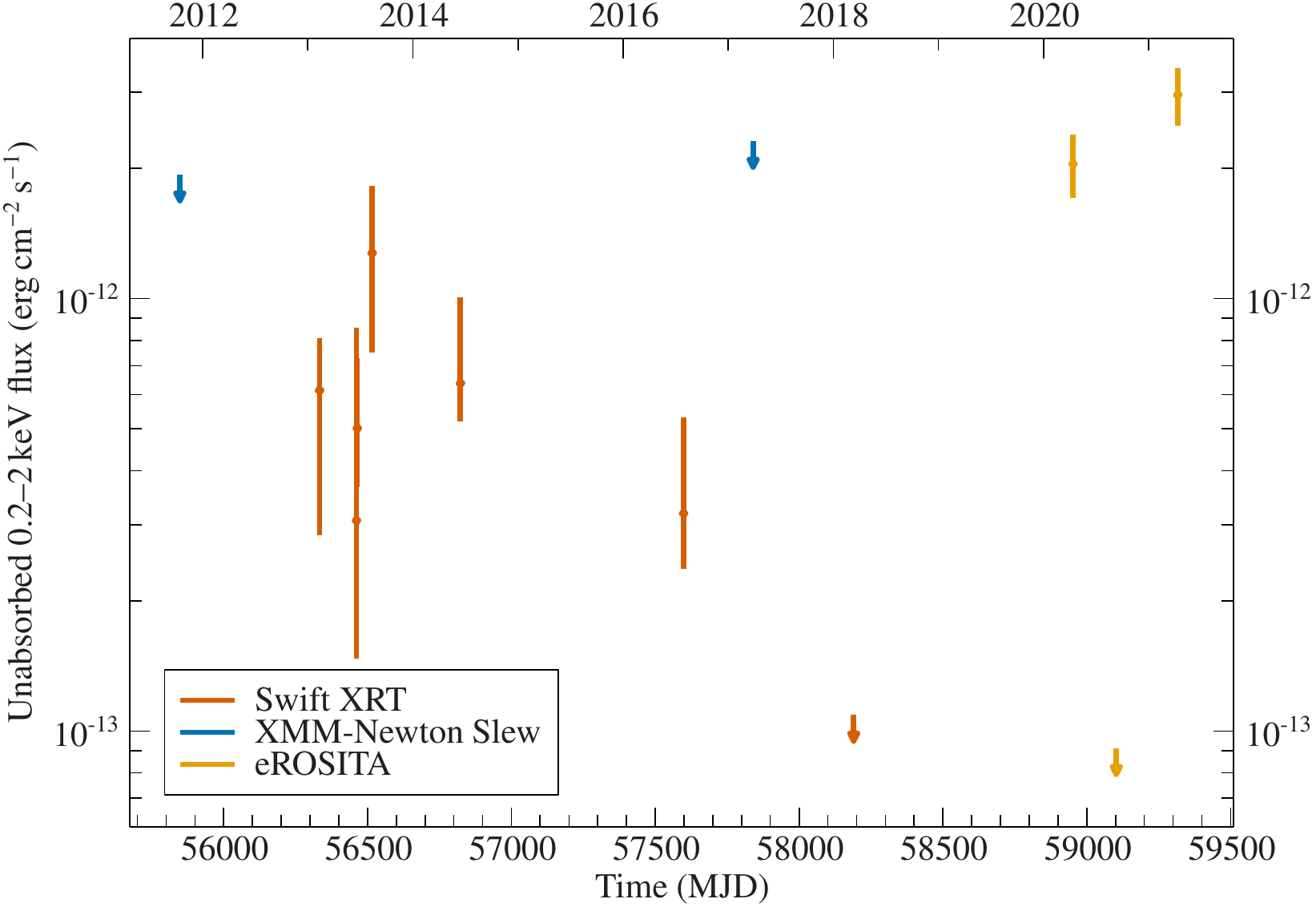}}
\caption{Long-term X-ray light curve of \att uisng \xmmn, \swi, and \ero data \label{f:lxlc}}
\end{figure}

\begin{figure}[t]
\resizebox{\hsize}{!}{\includegraphics{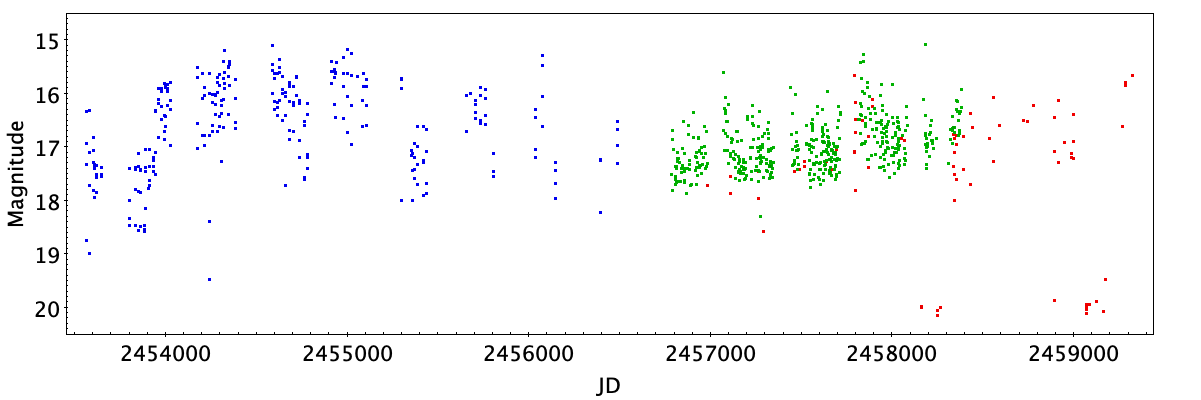}}
\caption{
CRTS (blue, white light), ASAS-SN (green, $g$-filter), and \gai (red, $G$)  archival photometric data. The magnitudes are shown as published.
\label{f:auxp}}
\end{figure}

The area of the transient was covered in eRASS1 between 2020-04-12 08:22:21 and 2020-04-13 04:22:41, in eRASS2 between 2020-10-14T20:59:40 and 2020-10-19 01:00:10 and in eRASS3 between 2021-04-11 21:22:14 and 2021-04-12 21:22:44 (all times in UTC). During those surveys 121 (228) and 181 (388) photons were registered with the standard source detection pipeline in the main band between 0.6 keV and 2.3 keV, respectively \citep[the numbers in parenthesis refer to the full band $0.2 -10$\,keV; all numbers given were derived with pipeline version c946, see][for details of the detection pipeline]{brunner+21}. The original photon event tables for each sky survey were binned to give mean count rates per scan\footnote{Individual scans are separated by 4 hours, the length of one eRODAY. Scans over a given celestial position may reach an exposure time of up to 40 s dependant on the off-axis angle}. Lightcurves in the full spectral bandpass, 0.2-10 keV, are shown in Fig.~\ref{f:elcs}. The mean rates in eRASS1 and 3 were 1.32 \rat and 1.92 \rat and the summed exposures 169 s and 223 s. The object showed up 100\% variability between erodays. In eRASS3 the fifth scan, centered on 2021-04-12 13:22:32 (UTC) lasting 37 seconds, revealed no source photon.

A joint X-ray spectral analysis of data obtained during eRASS1 and 3 was performed using the data from all seven telescope modules jointly. The pipeline processed spectra were grouped using the HEASARC {\tt grppha} command so that each bin contained a minimum of 16 photons. Given the CV-nature of the source a thermal plasma emission model was chosen initially, modified only by some cold interstellar absorbing matter ({\tt tbabs*APEC} in {\it XSPEC} terminology with {\it wilms} abundances). The data could well be represented by this model with $\chi^2 = 9.1$ for 15 degrees of freedom (Fig.~\ref{f:espc}). The fit does not imply strong constraints on the column density, $N_{\rm H}$ was found below $3\times 10^{20}$\,cm$^{-2}$ (90\% confidence), even compatible with zero. The main implication is that there is little if none intrinsic absorption in the system. The best-fit plasma temperature is at around 10\,keV ($>$5\,keV at 90\% confidence, not constrained towards high temperature). The bolometric flux during eRASS3 was $2.6 \times 10^{-12}$\,\fergs\ which gives a luminosity of $2.2\times 10^{31}$\,erg s$^{-1}$ \citep[assumed distance $376^{+15}_{-14}$\,pc,][]{bailer-jones+21}.

\section{Archival X-ray and optical photometric observations \label{s:xarc}}
There were several previous observations of the field of the new transient with the \swi satellite and with \xmmn \citep[slew survey,][]{saxton+08}. In Fig.~\ref{f:lxlc} all data are put in context. The source was not discovered with \xmmn i.e. was fainter than during eRASS1 or eRASS3. In the \swi data the source was detected 6 times, all data being consistent with each other at a flux of about $5 \times 10^{-13}$\,\fergs (0.2-2 keV). One \swi pointing revealed only a non-detection. In summary, over the past 10 years \att showed X-ray variability by a factor $>$20.

The new transient X-ray source was associated to the \gai object 6640165468504197120 at RA=19:29:33.110, DEC=$-$56:03:42.98, 2\farcs6 away from the X-ray position. This source was announced as \gai transient Gaia21bxo on March 14, 2021\footnote{\url{http://gsaweb.ast.cam.ac.uk/alerts/alert/Gaia21bxo/}, see also \url{https://www.wis-tns.org/object/AT2021kdp}}. \gai reports fading and re-brightening of the source by 4 mag with a floor at around $G \simeq 20.0$ and fuzzy maximum brightness. It was labeled as a cataclysmic variable (CV) candidate by the \gai team. Further archival photometric observations were reported by the CRTS and ASAS-SN automate sky surveys \citep{drake+09,shappee+14,kochanek+17} uncovering long- and short-term variability between 15th and 19th magnitude. The archival photometric data from CRTS, ASAS-SN and \gai are shown together in Fig.~\ref{f:auxp}. Pre-\gai data, in particular the CRTS, showed a high degree of optical variability on apparently long (years) and short (days) variability but the source was never seen before at magnitudes as faint as G=20.

\section{Optical follow-up observations\label{s:fup}} 
\subsection{SALT spectroscopy}
\begin{figure}[t]
\resizebox{\hsize}{!}{\includegraphics[clip=,angle=-90]{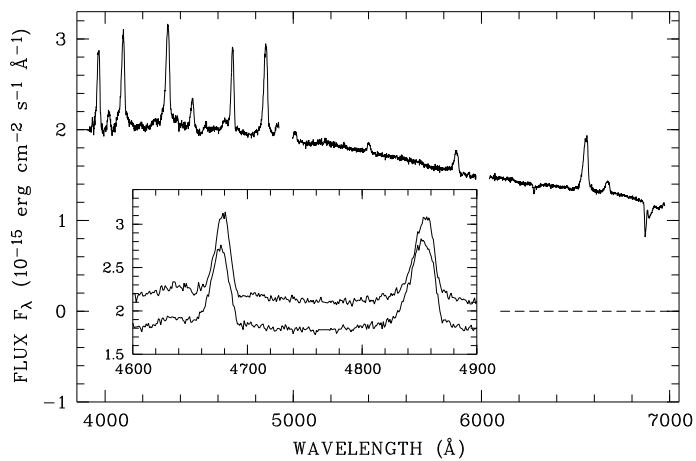}}
\caption{SALT spectra obtained of \att on 2021-05-06. The larger trace shows the mean of the two blue exposures while the inset shows both spectra individually zoomed on the spectral range with H$\beta$ and He{\sc II} $\lambda$4686. The dashed line in the main panel indicates zero spectral flux.\label{f:saltspec}}
\end{figure}

Two pairs of optical spectra, all 300 s exposure, were obtained using the RSS instrument \citep{Burgh2003} on the Southern African Large Telescope \citep[SALT;][]{Buckley2006}, beginning on 2021-05-06 23:53:42 UTC and finishing at 2021-05-07 00:17:10. The first pair of spectra covered the wavelength range 3920--6975\,\AA\ while the second pair of redder spectra covered 6035--9000\,\AA, both at a resolution of 5.5\AA. The blue continuum, which peaks at around 4500\,\AA, and pronounced emission lines of hydrogen and helium identify the object as a cataclysmic variable. The blue spectra are shown in Fig.~\ref{f:saltspec}. No significant features were seen in the red spectra other than the H$\alpha$ and HeI $\lambda$6678\AA\, emission lines.

The lines appeared rather symmetric at a blueshift of up to 600\,\kmps\ and with a FWHM corresponding to about 1500 \kmps. The measured equivalent width of H$\beta$ and of the ionized helium line He{\sc II} $\lambda$4686 was up to $\sim-$12 and $-8$\,\AA, respectively. The large flux ratio He{\sc II}/H$\beta \simeq 0.7$ suggests a magnetic nature, later confirmed from polarimetry.

\begin{table}
\caption{Optical and X-ray eclipse times of \att. Times are given as BJD in TDB. With \ero the eclipse length could not be measured.
\label{t:tecl}}
\begin{tabular}{rrll}
No & Cycle & Mid-eclipse time & Eclipse length\\
\hline
1 &   0 & 2459344.620060(17)& 0.003338 \\ 
2 &  15 & 2459345.583701(17)& 0.003338\\
3 &  16 & 2459345.647945(17)& 0.003335\\
4 &  30 & 2459346.547328(17)& 0.003333\\
5 &  31 & 2459346.611580(17)& 0.003334\\
6 &  32 & 2459346.675817(17)& 0.003328\\
7 & 107 & 2459351.494020(17)& 0.003326\\
8 & 109 & 2459351.622504(17)& 0.003332\\
9 & -429 & 2459317.0587(13)& --- \\
\hline
\end{tabular}
\end{table}

\subsection{HIPPO polarimetry and fast photometry}
Time resolved photopolarimetry using the two-channel all-Stokes polarimeter, HIPPO \citep{Potter2010}, was conducted on the SAAO 1.9-m telescope on 9, 10, 11 \& 16 May 2021. Observations were taken with clear, broad blue-band or broad red-band filters, with observations lasting from between 1.7 and 3.5 h on each night. 

\begin{figure}[t]
\resizebox{\hsize}{!}{\includegraphics{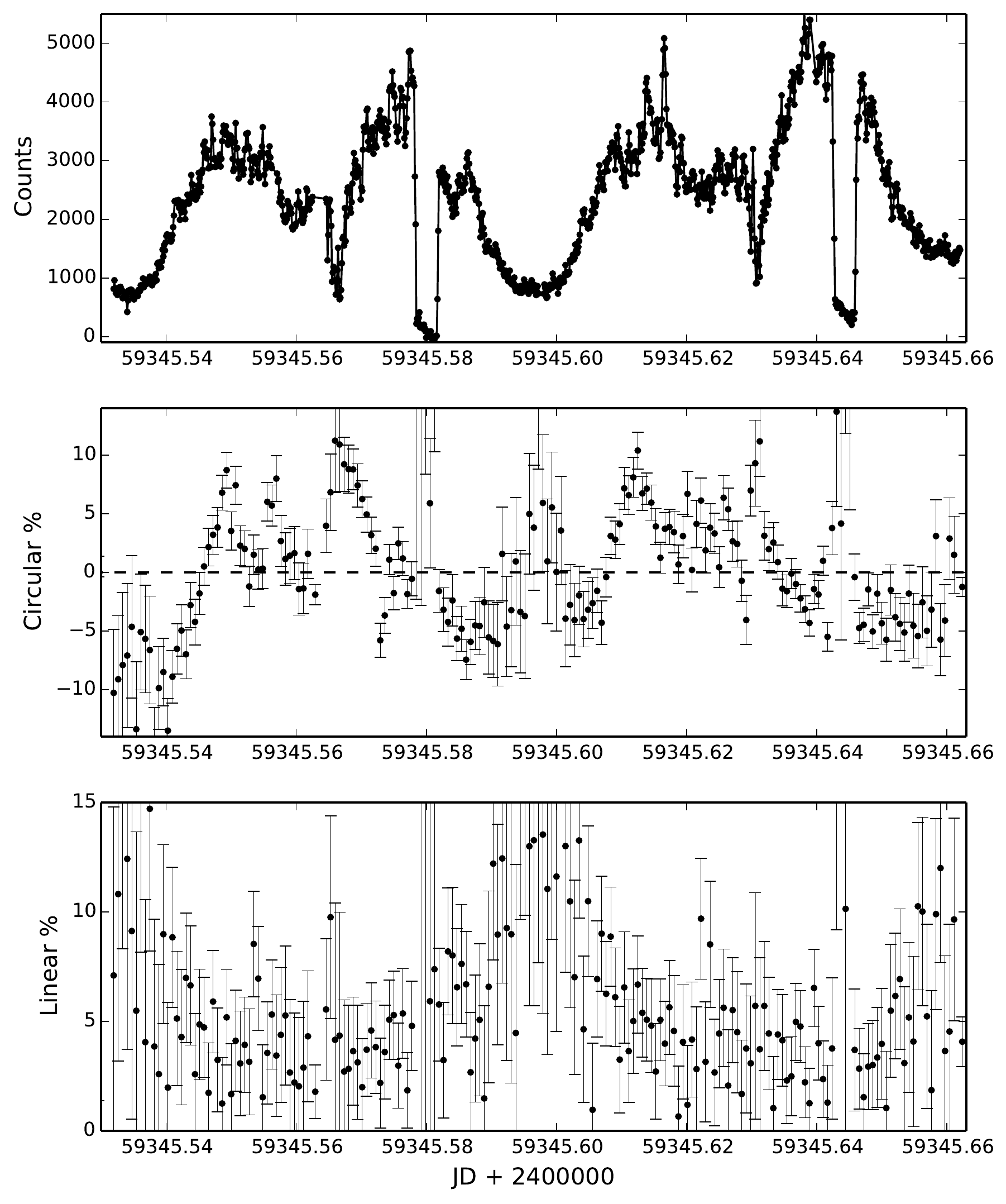}}
\caption{
HIPPO polarimetry on 10 May 2021. The upper panel shows the Stokes {\textit I} values, while the middle and bottom panels show the percentage circular and linear polarizations, respectively.
\label{f:pola}}
\end{figure}

HIPPO is a photon counting instrument with an intrinsic 1 ms time resolution, although the polarization data are accumulated every 0.1 s waveplate rotation cycle. The linear and circular  polarization, derived from the Stokes parameters, were derived after binning the data to 60 s time intervals. The results for the 10 May observations are presented in Fig.~\ref{f:pola}.
The circular polarization is seen to vary from -20 to 20\%, being higher in the blue, and linear polarization from 0 to 10\%, confirming the system to be a magnetic CV of the polar subclass. Detailed analysis and modelling of these observations will be the subject of a future paper.

\begin{figure}[t]
\resizebox{\hsize}{!}{\includegraphics{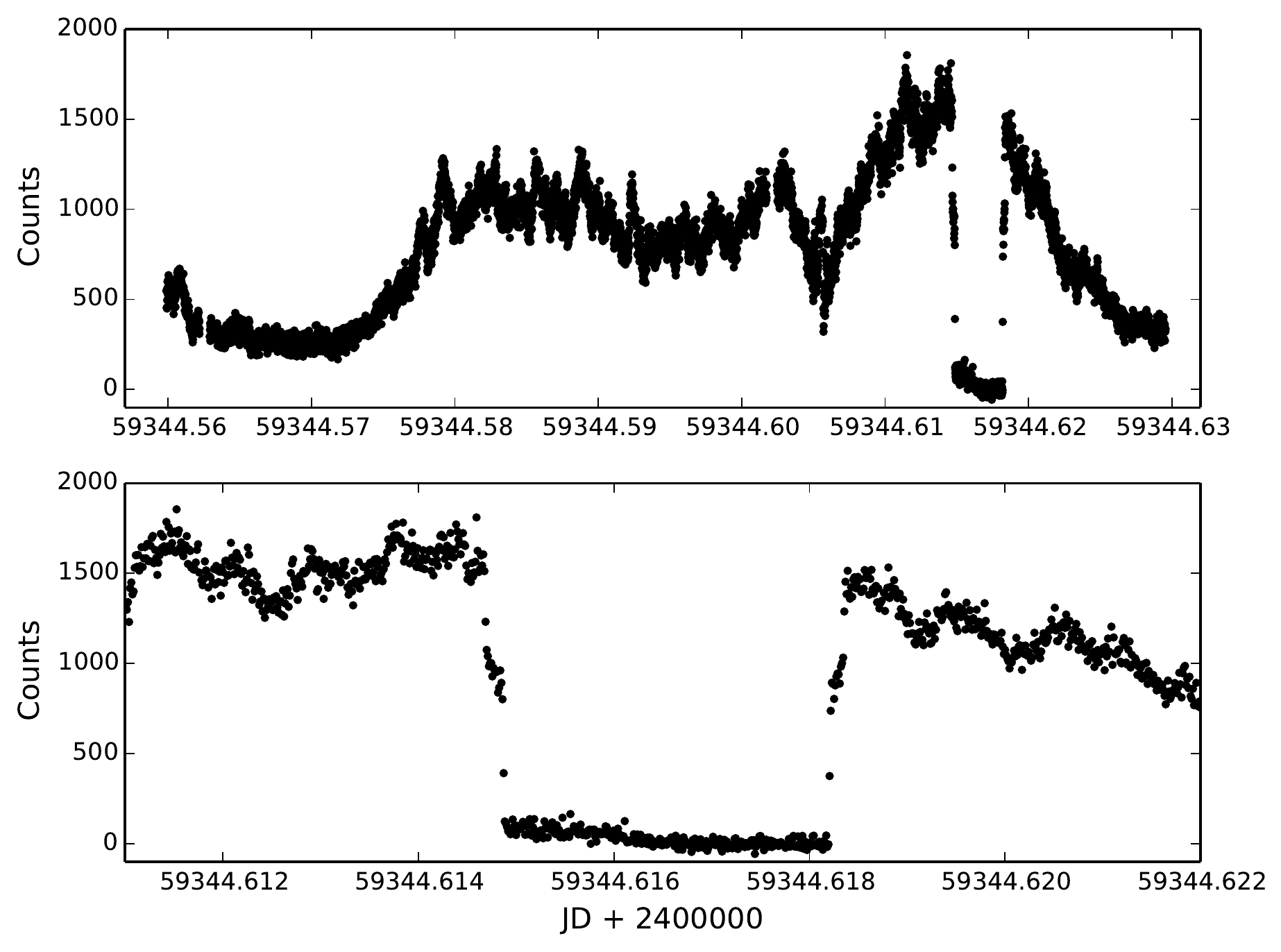}}
\caption{
Fast (1 s) photometry from HIPPO observations undertaken on 9 May 2021. Details of the eclipse are shown in the lower panel, in which the ingress and egress resolve the two accretion spots.
\label{f:phot}}
\end{figure}

Higher cadence (1\,s) photometry was derived from the HIPPO Stokes {\textit I} measurements. These revealed deep, structured eclipses, with ingress and egress showing two steps, each lasting roughly one second, indicating that the system is a two pole accretor (see Fig.~\ref{f:phot} for details of the 9 May observation). Eclipse timings were obtained from all four nights and listed in Table~\ref{t:tecl}.

For each eclipse the centers of the sharp second ingress and first egress, each lasting $\sim$2 s, was measured by eye, with the width of the ingress and egress defining the error. The eclipse length was estimated as the time between ingress (center) and egress (center). In reality the true eclipse length (defined as the flat bottom part of the eclipse) is $\sim$ 2 s shorter than the values shown below. The mean eclipse length was 288.0 s, with a standard deviation of 0.4 s. The eclipse lengths are also included in Table~\ref{t:tecl}.

The time taken between the end of the eclipse of the first hot spot and beginning of the eclipse of the second, is $\sim$15 s, i.e. the length of the first step at mid ingress. Likewise the length of egress step is $\sim$12 s.

We also note that the eclipses are not strictly flat bottomed, implying an additional source of luminosity is eclipsed after the white dwarf disappears, which is most likely to be the accretion stream. The depth of the eclipse, estimated from first to second contact (end of ingress), is $\sim$3 mag, while during eclipse the system declines by another $\sim$1 mag, at least. This implies that assuming a mean out of eclipse \textit{g} magnitude of 17 (from ASAS-SN), that the minimum brightness during eclipse is $\sim$21.

A linear regression among the optical data points listed in Table \ref{t:tecl} (first 8 entries) yielded the eclipse ephemeris as given in Eq.~\ref{e:eph}
\be
\mbox{BJD(TDB)} = 2459344.620058(9)+ E \times 0.0642426(2)
\label{e:eph}
\ee
where numbers in parenthesis give 1$\sigma$ uncertainties in the last digits. Hence, at an orbital period of $\sim$92\,min \att belongs to the short-period eclipsing objects. 

The X-ray non-detection during one of the eRASS3 scans suggests it coincided with one of the eclipses of the accreting white dwarf. This initial ephemeris was sufficiently accurate to extend back by one month and determine the cycle number with high accuracy. The corresponding time is given in the last row of Table \ref{t:tecl}. The timing error was assumed as eclipse half length minus length of the eroday (40sec) to stay completely within the eclipse. A linear regression among all data in the table yields leaves all values and their errors unchanged within the error margins of Eq.~\ref{e:eph} which thus represents our final result.

The \gai light curve has a floor at $G\sim20 $ (see the red data points in Fig.~\ref{f:auxp}) and we were tempted to associate these times with further past eclipse observations. The group of the more recent \gai events lies 170-270 days before our time zero in Eq.~\ref{e:eph}. The accumulated timing uncertainty over an interval of 250 days is 51 seconds (0.009 phase units). None of the \gai data points lies close to predicted phase zero. We conclude that the low \gai values are not eclipse events.

\subsection{Results and conclusion}
We report the discovery of the bright, strongly variable object \att, which was detected in the ongoing \ero X-ray all-sky surveys. It was identified as a magnetic CV, belonging to the polar subclass, through optical and X-ray spectroscopy, high-speed optical photometry and polarimetry. While it is expected that the eRASS surveys will uncover many such objects, this one is special through its high brightness and its structured eclipse discovered by our initial high-speed polarimetry and photometry. The deep, $\sim$4 mag, eclipse lasts about 300 sec and allowed to derive a precise binary orbital period of $92.5094\pm0.0002$\,min. The non-detection of the source during one of the eRASS3 scans was due to a coincidence with the eclipse of the accreting white dwarf. 

The eclipses themselves display a very interesting two-step structure (lower panel of Fig.~\ref{f:phot}) similar to that of UZ For \citep{perryman+01} indicating accretion at two poles. Only a small phase offset was observed, hence it appears likely that there is one 'northern' and one 'southern' pole at perhaps similar stellar longitude (observer located in the 'northern' hemisphere). The northern one (shorter eclipse) was found about 70\% brighter in the $I$-band than the southern one. The bright phase is centred at phase $\sim$0.75. A pre-eclipse dip occurrs and is centred at phase $\sim$0.2, but begins already at phase $\sim$0.77, i.e.~roughly 80 degrees before eclipse center. The pre-eclipse marks the occasion when our line-of sight crosses the stream feeding the northern pole. The early phase of this feature and the centroid of the optical lightcurves point to a somewhat unusual location of the northern, presumably the main, accretion spot roughly a quarter cycle before eclipse.  The stellar longitude of the southern spot is uncertain right now, but it must lie closer to the binary meridian (line connecting both stars) because its eclipse is centred slightly earlier in phase than that of the northern region (as indicated by the different lengths of the slow ingress/egress phases between the steep ingress/egress phases). 

Interestingly, \att becomes strongly positively polarized through the pre-eclipse dip. If the light from the northern spot is completely shielded at the dip, this must be synchrotron radiation from the southern spot originating at a different longitude. 

Circular polarimetry shows several zero crossings but no phase of complete absence of polarization. There must be always one of the spots in the field of view of the observer. Even during the extended phase of minimum brightness of the light curve, at around phase 0.25, a linear polarization pulse reaching almost 15\% and some positive circular polarization are found indicating the visibility of at least one of the accretion regions. 

The SALT spectra were obtained in the phase interval between $0.42 - 0.67$. Their blue continuum, if interpreted as high-harmonic cyclotron radiation forming a quasi-continuum is indicative of a moderate field strength of about 20-30\,MG \citep[see e.g.][for cyclotron spectra of polars in that range of field strength]{schwope_beuermann90,schwope+93}. They expectedly do not show any sign of the secondary star. If the donor star in \att would follow the revised sequence of \citet{knigge07} for normal CVs it would have a mass 0.108\,\msun (spectral type about M6.5), an absolute magnitude $M_{V} = 14.9$, and an apparent magnitude $V=22.8$ for the \gai-based distance, too faint to be detected with our spectroscopy or fast-photometry at the bottom of the eclipse.

For an estimate of the orbital inclination we follow the method outlined by \citet{chanan+76}. For an  assumed eclipse length of the white dwarf of 302 seconds, and a mass ratio $M_{\rm WD}/M_2=0.8/0.108=7.4$ one gets $i=82.5\degr$. For an extreme mass ratio of 10 the maximum inclination is $84.4\degr$.


\att was not detected with \ros with the likely explanation of a low state at those days. If it would have been in a high state during the RASS, it nevertheless would probably have escaped its identification. It does not show the pronounced soft X-ray component, triggering follow-up of soft sources that led to the discovery of many polars from the RASS \citep[e.g.][]{beuermann+schwope94}. It is located at a galactic latitude of $-27.4\degr$. The Galactic Plane Survey \citep[GPS,][]{motch+96} observed below $|b^{II}| = 20\degr$, while the systematic follow-up of all bright sources, the ROSAT Bright Survey \citep[RBS,][]{schwope+00,schwope+02} observed only above $|b^{II}| = 30\degr$. Hence, there is a huge discovery space to be explored with the systematic follow-up of \ero sources that will detect objects like \att out to a discance of 1500\,pc.

\begin{acknowledgements}

This work is based on data from \ero, the soft instrument aboard SRG, a joint Russian-German science mission supported by the Russian Space Agency (Roskosmos), in the interests of the Russian Academy of Sciences represented by its Space Research Institute (IKI), and the Deutsches Zentrum für Luft- und Raumfahrt (DLR). The SRG spacecraft was built by Lavochkin Association (NPOL) and its subcontractors, and is operated by NPOL with support from the Max Planck Institute for Extraterrestrial Physics (MPE).

The development and construction of the \ero X-ray instrument was led by MPE, with contributions from the Dr. Karl Remeis Observatory Bamberg \& ECAP (FAU Erlangen-Nuernberg), the University of Hamburg Observatory, the Leibniz Institute for Astrophysics Potsdam (AIP), and the Institute for Astronomy and Astrophysics of the University of Tübingen, with the support of DLR and the Max Planck Society. The Argelander Institute for Astronomy of the University of Bonn and the Ludwig Maximilians Universität Munich also participated in the science preparation for \ero.

The \ero  data shown here were processed using the eSASS/NRTA software system developed by the German \ero consortium.

Some of the observations presented here were obtained with SALT under the transients followup programme 2018-2-LSP-001 (PI: DB), which is supported by Poland under grant no. MNiSW DIR/WK/2016/07. DB and SP also acknowledge research support from the National Research Foundation.

MG is supported by the EU Horizon 2020 research and innovation programme under grant agreement No 101004719.

\end{acknowledgements}

\bibliography{at20}

\end{document}